\documentclass{iau}
\usepackage[dvips]{graphicx}

\usepackage{natbib}

\newcommand{\aap}{\textit{A\&A}}
\newcommand{\aaps}{\textit{A\&AS}}
\newcommand{\apj}{\textit{ApJ}}
\newcommand{\apjl}{\textit{\apj}}

\newcommand{\aj}{\textit{AJ}}
\newcommand{\mnras}{\textit{MNRAS}}

\newcommand{\kmps}{\mathrm{km~s^{-1}}}
\newcommand{\ion}[2]{#1$\,${\sc {#2}}}   

\title[Magnetospheric Accretions and the Inner Winds of Classical T
  Tauri Stars] 
{Magnetospheric Accretions and the Inner Winds of Classical T Tauri Stars}

\author[Ryuichi Kurosawa \& M. M. Romanova]   
{Ryuichi Kurosawa$^1$
 \and M. M. Romanova$^2$}

\affiliation{$^1$ Max-Planck-Institut f\"{u}r Radioastronomie, \\ 
                  Auf dem H\"{u}gel, 69, 53212 Bonn, Germany \\ 
                  email: {\tt kurosawa@mpifr-bonn.mpg.de} \\[\affilskip]
             $^2$ Department of Astronomy, Cornell University, \\ 
                  Ithaca, NY 14853-6801, USA \\
                  email: {\tt romanova@astro.cornell.edu}}

\pubyear{2013}
\volume{302}  
\pagerange{119--126}
\jname{Magnetic Fields Throughout Stellar Evolution}
\editors{A.C. Editor, B.D. Editor \& C.E. Editor, eds.}
\begin{document}

\maketitle

\begin{abstract}
Recent spectropolarimetric observations suggest that young low-mass
stars such as classical T Tauri stars (CTTSs) possess relatively
strong ($\sim$kG) magnetic field. This supports a scenario in which
the final accretion onto the stellar surface proceeds through a
magnetosphere, and the winds are formed in magnetohydrodynamics (MHD)
processes.  We examine recent numerical simulations of magnetospheric
accretions via an inclined dipole and a complex magnetic
fields. The difference between a stable accretion regime, in which
accretion occurs in ordered funnel streams, and an unstable regime, in
which gas penetrates through the magnetosphere in several unstable
streams due to the magnetic Rayleigh-Taylor instability, will be
discussed. We describe how MHD simulation results can be used in
separate radiative transfer (RT) models to predict observable
quantiles such as line profiles and light curves. The plausibility of
the accretion flows and outflows predicted by MHD simulations (via RT
models) can be tested against observations. We also address the issue of
outflows/winds that arise from the innermost part of CTTSs. First, we
discuss the line formations in a simple disk wind and a stellar wind
models. We then discuss the formation of the conically shaped
magnetically driven outflow that arises from the disk-magnetosphere
boundary when the magnetosphere is compressed into an X-type
configuration.


\end{abstract}

\firstsection 

\section{Introduction}
\label{sec:intro}

In the standard picture of accretions in classical T~Tauri stars
(CTTSs), the stellar magnetic fields are assumed to be strong enough
to truncate the accretion disks near the star-disk corotation radii
(e.g., \citealt{Camenzind:1990}; \citealt{Koenigl:1991}).  The matter
originating in the vicinity of the disk truncation radius flows along
the magnetic field lines to the stellar surface.  The kinetic energy
of the infalling matter is converted into the thermal energy of the
plasma in the shock regions above the photosphere.  This general
picture of magnetospheric accretions successfully explains some of
the important observational aspects of CTTSs, e.g., the UV-optical
continuum excess (e.g.~\citealt{Calvet:1998}) and the inverse P-Cygni
profiles (e.g.~\citealt{hartmann94}; \citealt{Muzerolle:2001}). In the
earlier magnetospheric accretion models (e.g.~\citealt{Koenigl:1991};
\citealt{hartmann94}), the flows are often assumed to be steady and
axisymmetric, as in the original models for accreting neutron stars
(e.g.~\citealt{Ghosh:1977}) . While these axisymmetric models agree
with some of the observational phenomena as mentioned above, they are
not applicable to many other observational aspects.  For example,
(1)~some CTTSs are known to exhibit complex line and continuum
variability (e.g.~\citealt{Johns:1995};
\citealt{Petrov:1996}), and (2)~recent time-series spectropolarimetric
(Zeeman-Doppler Imaging) observations
(\citealt{Donati:2007,donati11,Jardine:2008, Gregory:2008,
  Hussain:2009}) have revealed that some of CTTSs have complex stellar
magnetic fields.  Clearly, these observational aspects (complex
variability and accretion geometry) cannot be explained by simple
axisymmetric models.  To address the issues of the complex variablity
and non-axisymmetry of CTTSs, we will use three-dimensional
(3-D) radiative transfer models of time-series line profiles using the
results of  3-D magnetohydrodynamics (MHD) 
simulations of the accretion onto CTTSs through inclined magnetic
multipoles.  The results are presented in Sec.~\ref{sec:MA}

Not only the accretion flows, but also the outflows/winds in CTTSs are
most likely shaped by magnetic fields. The formation of the outflow
itself likely occurs in MHD processes in which open magnetic fields
are anchored to a star, an accretion disk or both (e.g.,
\citealt{Ferreira:2006}). Observationally, there are ample of
spectroscopic evidences that rather high velocity winds/outflows
originate from the innermost part of the star-disk systems, e.g., the
blueshifted absorption component in strong optical and 
near-UV lines such as~ H$\alpha$, \ion{Na}{i}~D, \ion{Ca}{ii} H\&K
(e.g., \citealt{Reipurth:1996, Alencar:2000, Ardila:2002}). In addition, the
near-infrared \ion{He}{i}~$\lambda$10830 line has been recognized as a
robust wind diagnostic line by e.g.\ \citet{Takami:2002},
\citet{Edwards:2003}, \citet{Edwards:2006}. In particular,
\citet{Edwards:2006} showed that the line is very sensitive to the
presence of the wind, and about 70~per~cent of CTTSs exhibit a
blueshifted absorption (below continuum) in \ion{He}{i}~$\lambda$10830
while only about 10~per~cent of CTTSs show a similar type of
absorption component in H$\alpha$ (\citealt{Reipurth:1996}).
\citet{Edwards:2006} and \citet{Kwan:2007} suggested that the
blueshifted absorption component in \ion{He}{i}~$\lambda10830$
profiles is cased by a stellar wind in about 40~per~cent, and by a
disc wind in about 30~per~cent of the samples in \citet{Edwards:2006}.
In Sec.~\ref{sec:wind}, we explore how the radiative transfer models
of \ion{He}{i}~$\lambda$10830 can be used to probe the origin of the
inner winds of CTTSs.  We will also test the plausibility of the
conically shaped magnetically driven wind (``the conical wind'')
solution found in the MHD simulations of \citet{Romanova:2009}.

\section{Non-Axisymmetric Accretion Models and Complex Variablity}
\label{sec:MA}

In this section, we examine the types of variability expected from 
(1)~\emph{steady} non-axisymmetric accretion flows
(Sec.~\ref{subsec:V2129Oph}), and (2)~\emph{unsteady} non-axisymmetric
accretion flows (Sec.~\ref{subsec:instab}).

\subsection{Rotationally Induced Modulations Caused by Steady
  Accretion Flows} 
\label{subsec:V2129Oph}

\begin{figure*}
  \begin{center}
      \includegraphics[clip,width=1.0\textwidth]{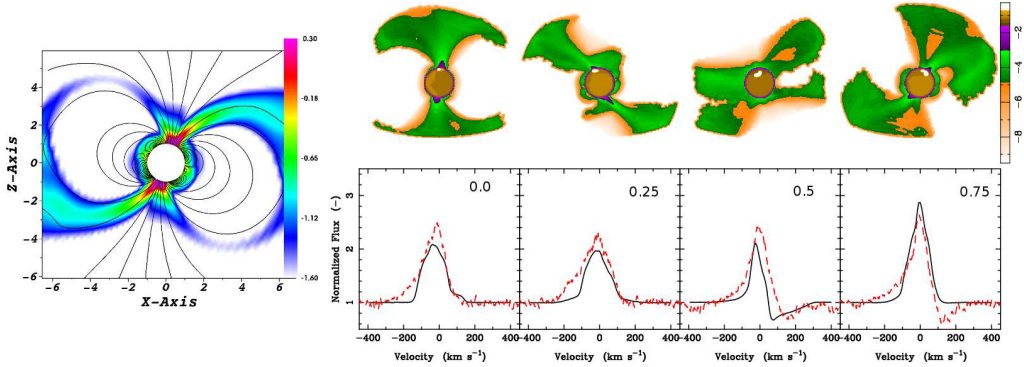} \hspace{1.3cm} 
  \end{center}
\vspace{0.0cm}
\caption{\emph{Left panel}: A cross section of the density from the
  3-D MHD simulations of accretion onto the CTTS V2129~Oph. The
  density is shown in logarithmic scale and in arbitrary units. The
  solid lines represent sample magnetic field (a combination of a
  dipole and an octupole).  \emph{Right panels}: The radiative transfer
  models computed for the accretion flows from the MHD
  simulation. \emph{Upper panels} show the H$\delta$ emission maps at
  four different rotation phases of the star (0.0, 0.25, 0.5 and 0.7
  from left to right). \emph{Lower panels} show the corresponding
  line profiles models of H$\delta$ (solid) along with the data obtained by
  ESPaDOnS (dashed). The models and data are from \cite{alencar12}.
 }
\label{v2129oph}
\end{figure*}

A set of radiative transfer models which simulate the rotationally
induced line variability arising from a complex circumstellar
environment of CTTSs was presented by \citet{Kurosawa08} (see also
\citealt{Symington:2005, Kurosawa:2005}). The results of the 3-D MHD
simulations from \citet{Romanova:2003} and \citet{Romanova:2004}, who
considered accretion onto a CTTS with a misaligned dipole magnetic
axis with respect to the rotational axis, were used in a separate 3-D
radiative transfer model (e.g., \citealt{harries2000, Kurosawa:2006})
to predict observed line profiles (e.g., H$\beta$, Pa$\beta$ and
Br$\gamma$) as a function of rotational phases.  The accretion in
those MHD models are in a steady state, and the accretion mainly
proceeds in two funnel flows, i.e., one lands on the upper and the
other on the lower hemispheres. In this study, the general
dependencies of line variability on inclination angles ($i$) and
magnetic axis misalignment angles ($\Theta$) were examined.

Using the same method, we now examine the accretion flows with more
complex magnetic field configurations (with a combination of an
inclined dipole and an octupole components), similar to those in
\citet{Romanova:2011} and \citet{Long:2011}.  
In particular, the accretion flows around the CTTS 
V2129~Oph (K5) were modeled by the MHD simulations in which the
information from the surface magnetic field map (by the Zeeman-Doppler
imaging technique) obtained by \citet{donati11} was incorporated (see
\citealt{alencar12}).  For example, the observation by
\citet{donati11} suggests that the inclinations of the magnetic dipole
and the octupole are about $15^{\circ}$ and $25^{\circ}$ with
respective to the rotation axis, and the octupole axis is about 0.1
ahead of the dipole in the rotational phase. A single time-slice of the
MHD simulations was used to predict the dependency of the observed
line profiles on the rotational phase by using the 3-D radiative transfer model
mentioned earlier.  The results are shown in Fig.~\ref{v2129oph}.  As
one can see from this figure, near the stellar surface, the octupole
component dominates and it redirect the accretion funnels to a higher
latitude. On the other hand, the dipole component dominates in larger scales,
and it interact with the innermost part of the accretion disk.  
The line profile models for H$\beta$ are compared with the observations
obtained by ESPaDOnS at CFHT (\citealt{alencar12}) at four different rotational
phases (0.0, 0.25, 0.5 and 0.75). The figure shows that our model
agrees with the time-series observed line profiles very well.  More
detailed comparisons of the model with observations can be found in
\citet{alencar12}.

\subsection{Irregular Variability Caused by Accretion in an Unstable Regime}
\label{subsec:instab}

\begin{figure*}
  \begin{center}
    \begin{tabular}{cc}
      \includegraphics[clip,width=0.4\textwidth]{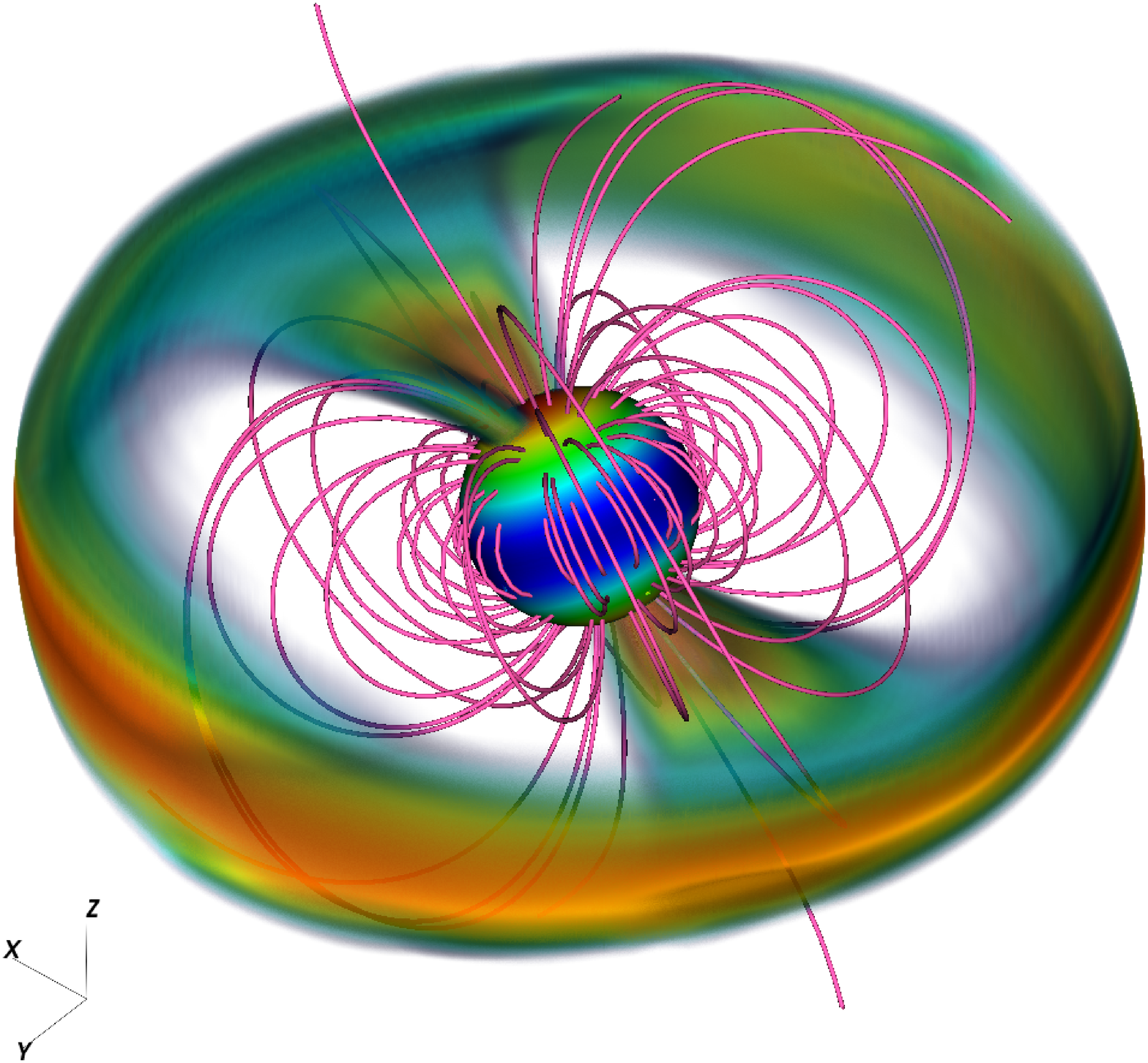}&
      \includegraphics[clip,width=0.5\textwidth]{kurosawa.ryuichi_fig02b.eps}\\     
      \includegraphics[clip,width=0.4\textwidth]{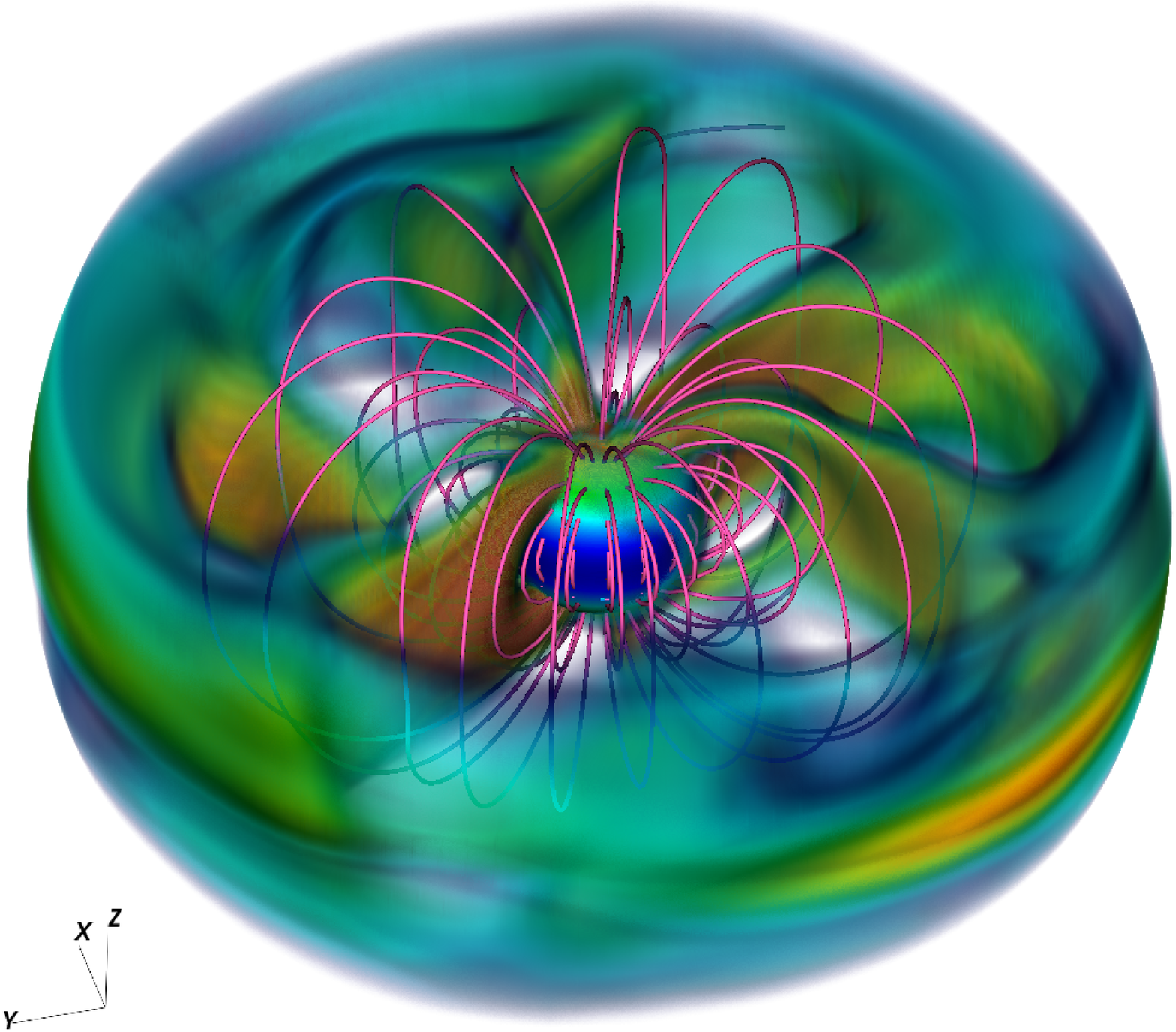}&
      \includegraphics[clip,width=0.5\textwidth]{kurosawa.ryuichi_fig02d.eps}\\  
    \end{tabular}
  \end{center}
\vspace{-0.1cm}
\caption{A comparison of accretions onto a CTTS in a stable (\emph{upper panels}) and an unstable
  (\emph{lower panes}) regimes. While the density distributions (volume renderings; in
  logarithmic scales; in arbitrary units) from the 3D MHD simulations
  are shown in the \emph{left panels}, the radiative transfer
  models for the continuum flux and  H$\delta$ line
  equivalent widths plotted as a function of rotation phases are shown
  on the \emph{right panels}.  The continuum flux are evaluated near
  H$\delta$.  The figures are from \citet{kurosawa13}. 
}
\label{rt-instab}
\end{figure*}

In the previous section (Sec.~\ref{subsec:V2129Oph}), we 
considered the cases in which the accretion flows are steady even
though the flows themselves are non-axisymmetric. Here, we examine the
unsteady accretion flows caused by the magnetic Rayleigh-Taylor (R-T)
instability, which occurs at the interface between an accretion disk
and a stellar magnetosphere. The instability tends to appear in a
system with a relatively slowly rotating magnetosphere, a high
mass-accretion rate and a small misalignment angle for the magnetic poles.
The detail theoretical studies on the conditions for instability have
been presented by e.g., \cite{arons76,spruit93,li:2004}.
More recently, the global three-dimensional (3D) magnetohydrodynamic
(MHD) simulations by e.g., \cite{roma08,kulk08,kulk09} have shown that
the R-T instability induces accretions in multiple (a few to several)
unstable vertically elongated streams or `tongues' which penetrate 
the magnetosphere (the lower left panel in Fig.~\ref{rt-instab}). They
found that the corresponding time-scale of the 
variability induced by the instability is typically a few times
smaller than the rotation period of a star.  Here, we examine the
fundamental differences in the observational properties of the
accretions in the stable and unstable regimes.

The results of global 3-D MHD simulations of matter flows in both
stable and unstable accretion regimes are used in \citet{kurosawa13}
to calculate time-dependent hydrogen line profiles and study their
variability behaviors (see Fig.~\ref{rt-instab}).  In the stable
regime, some hydrogen lines (e.g., H$\beta$, H$\gamma$, H$\delta$,
Pa$\beta$ and Br$\gamma$) show a redshifted absorption component only
during a fraction of a stellar rotation period, and its occurrence is
periodic.  However, in the unstable regime, the redshifted absorption
component is present rather persistently during a whole stellar
rotation cycle, and its strength varies non-periodically.  In the
stable regime, an ordered accretion funnel stream passes across the
line of sight to an observer only once per stellar rotation period
while in the unstable regime, several accreting streams/tongues, which
are formed randomly, pass across the line of sight to an observer.
The latter results in the quasi-stationarity appearance of the
redshifted absorption despite the strongly unstable nature of the
accretion.  In the unstable regime, multiple hot spots form on the
surface of the star, producing the stochastic light curve with several
peaks per rotation period.  Interestingly, such irregular light-curves
are frequently observed in CTTSs
(e.g.~\citealt{herbst94,rucin08,alencar10}).  Note that irregular
light curves are found in about 39~per~cent of the CTTS samples in
\cite{alencar10}.  No clear periodicity in the line variability is
found in many CTTSs. This study suggests a 
CTTS that exhibits a stochastic light curve and a stochastic line
variability, with a rather persistent redshifted absorption component,
may be accreting in the unstable accretion regime.

\section{The Inner Winds of Classical T Tauri Stars}
\label{sec:wind}

Understanding the origin of an outflow, whether it is a stellar wind,
the X-wind (\citealt{Shu:1994}), the conical wind
(\citealt{Romanova:2009}) or disk wind, 
is important as it is closely related to the angular momentum evolution
of young stellar objects. Here, we demonstrate how the line profile
models can be used to probe the origin of the inner winds. 

\subsection{Simple Kinematics Wind Models}
\label{subsec:simple-wind}

\begin{figure*}
  \begin{center}
      \includegraphics[clip,width=0.95\textwidth]{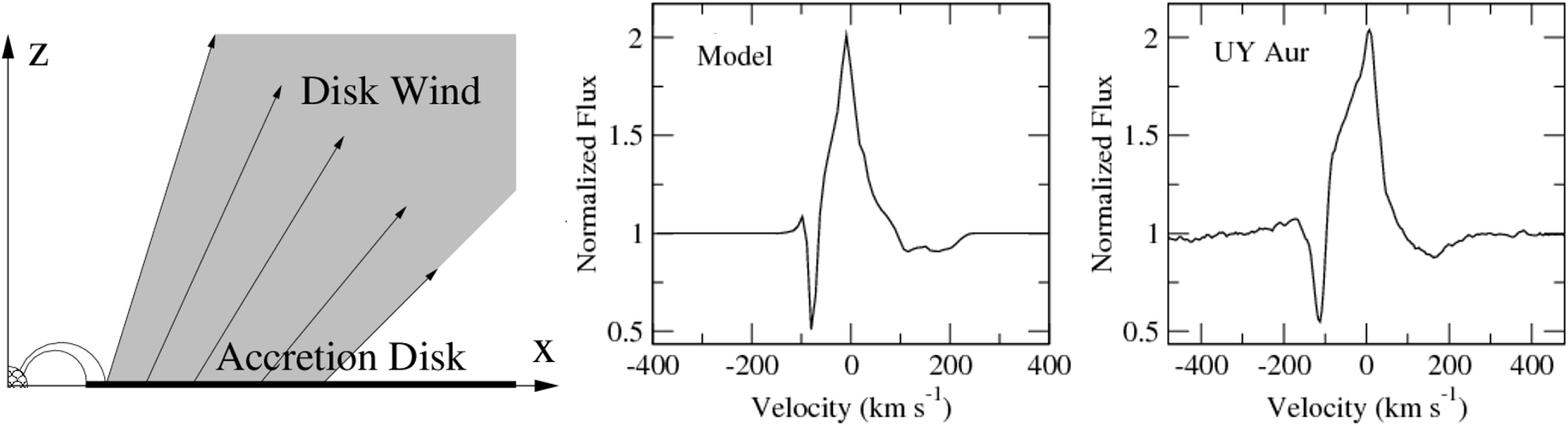} \vspace{0.5cm} \\
      \includegraphics[clip,width=0.95\textwidth]{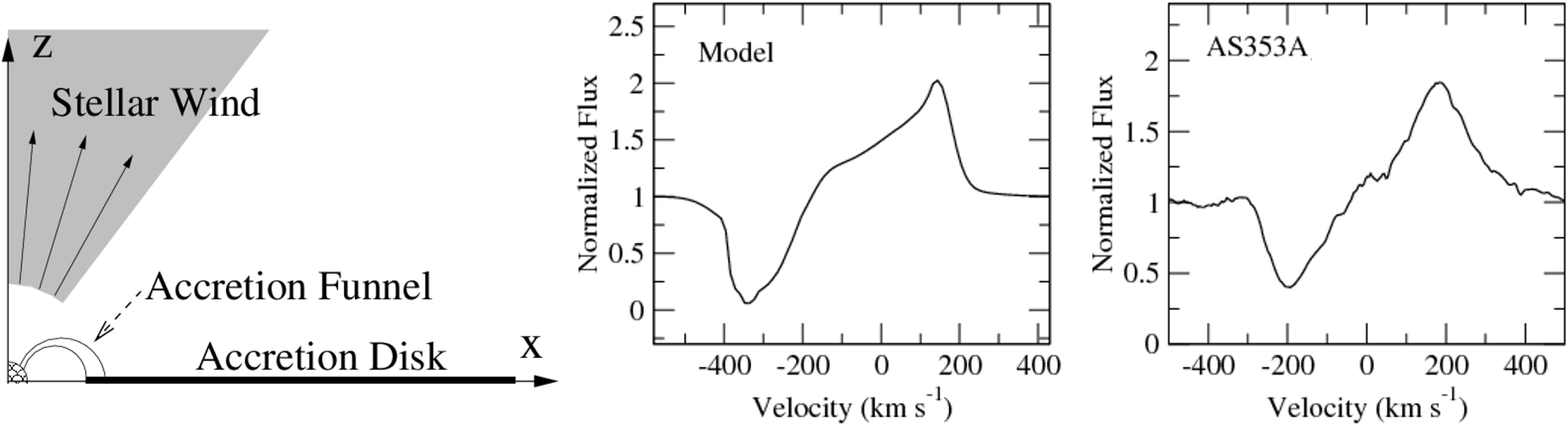}
  \end{center}
\vspace{0.0cm}
\caption{Simple axisymmetric kinematic disk wind (\emph{upper panels}) and stellar
  wind (\emph{lower panels}) models that are combined a dipolar
  magnetospheric accretion. The \emph{left panels} show the flow
  configurations.  The corresponding line profile models of
  \ion{He}{i}~$\lambda$10830 are shown in the \emph{middle
    panels}. The inclination angles ($i$) used for computing the line
  profiles are $i=48^{\circ}$ and $10^{\circ}$ for the models with a disk wind and
   a stellar wind, respectively.  The observed \ion{He}{i}~$\lambda$10830 line profiles
  for the CTTSs UY~Aur and AS~353~A (from \citealt{Edwards:2006}) 
  are shown in the \emph{right panels}, for qualitative
  comparisons. The figures are from \cite{kurosawa11}. 
 }
\label{simple-wind}
\end{figure*}

As briefly mentioned in Sec.~\ref{sec:intro}, \citet{Edwards:2006}
have demonstrated a robustness of the near-infrared
\ion{He}{i}~$\lambda$10830 line for probing the inner winds of
CTTSs. In particular, they have classified 
\ion{He}{i}~$\lambda$10830 line profiles into two types based on the
shapes of the blueshifted wind absorption component. The first type
has a relatively narrow and low wind absorption component (the upper
right panel in Fig.~\ref{simple-wind}). The second type has a wide and
deep wind (P-Cygni like) absorption component which reaches the
maximum velocity of 300--400\,$\kmps$ (the lower right panel in
Fig.~\ref{simple-wind}).  \citet{Edwards:2006} suggested that the
former is caused by a disk wind, and the star is viewed 
at a relatively high inclination angle at which a line of sight can
intersect with the disk wind. On the other hand, the latter is cased
by a stellar wind, and the star is viewed nearly pole-on.  

To test the scenarios of \citet{Edwards:2006} for the two distinctive
wind absorption features, we have constructed the simple kinematic
flow models around CTTSs. See \citet{kurosawa11} for the model
descriptions. The first model consists of a disk wind and
magnetospheric accretion funnels (the upper left panel in
Fig.~\ref{simple-wind}). The second model consists of a stellar wind
and magnetospheric accretion funnels (the lower left panel in
Fig.~\ref{simple-wind}). The corresponding line profile models for
\ion{He}{i}~$\lambda$10830 are also shown in the same figure (the
middle panels). Our disk wind + magnetosphere model reproduces not
only the narrow blueshifted wind absorption component, but also the
redshifted absorption component as seen in the observed
\ion{He}{i}~$\lambda$10830 line profile of the CTTS UY~Aur.  Note that
the redshifted absorption component is caused by the infalling gas in
the magnetospheric accretion funnel. Similarly, our stellar wind +
magnetosphere model reproduces a rather wide and deep blueshifted
absorption component as seen in the observed
\ion{He}{i}~$\lambda$10830 in the CTTS AS~353~A.  Interestingly, the
narrow wind absorption caused by the disk wind is present in the line
profile only when the inclination angle of the system is an
intermediate to a high value because of the geometry of the disk
wind. Similarly, the wide and deep P-Cygni like wind absorption is
present only at a very low inclination angle (near pol-on).  Hence,
our simple kinematic wind models confirms the earlier finding of
\citet{Edwards:2006} who suggested two different types of wind for the
narrow and wide blueshifted absorption components seen in
\ion{He}{i}~$\lambda$10830.  A similar analysis was also performed by
\citet{Kwan:2007} with simplified line emissivity and opacity in the
winds.

In reality (as also suggested by \citealt{Edwards:2006}), both types
of the winds (a stellar and a disk winds) could coexist; however, it
is likely that only one type of wind absorption component appears in a
line profile because the line of sight to the stellar surface could
intersect only one type of wind for a given viewing angle of the
system.

\subsection{The Conical Wind Model}
\label{subsec:conical-wind}

\begin{figure*}
  \begin{center}
    \includegraphics[clip,width=0.95\textwidth]{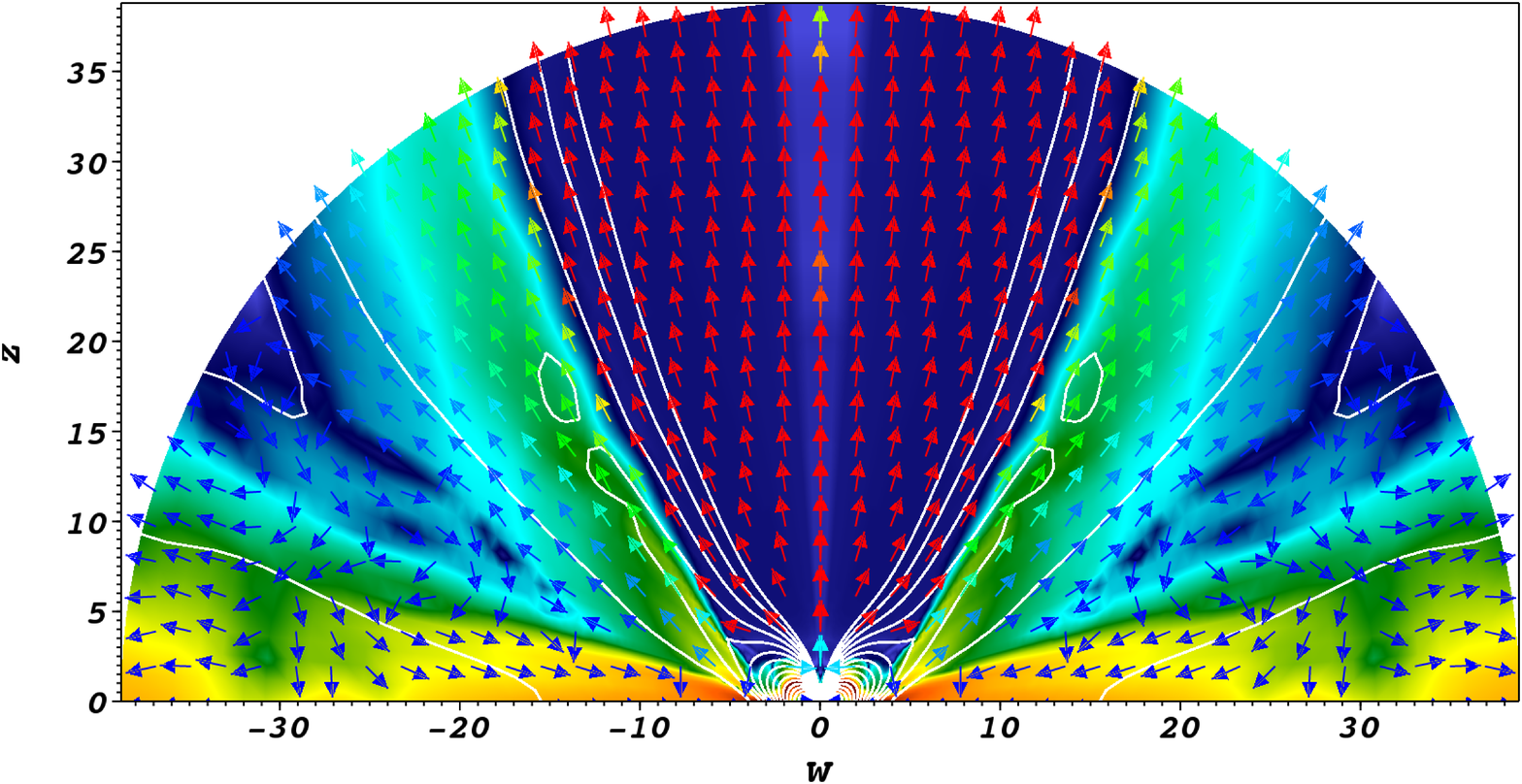} \vspace{0.3cm} \\
    \includegraphics[clip,width=0.6\textwidth]{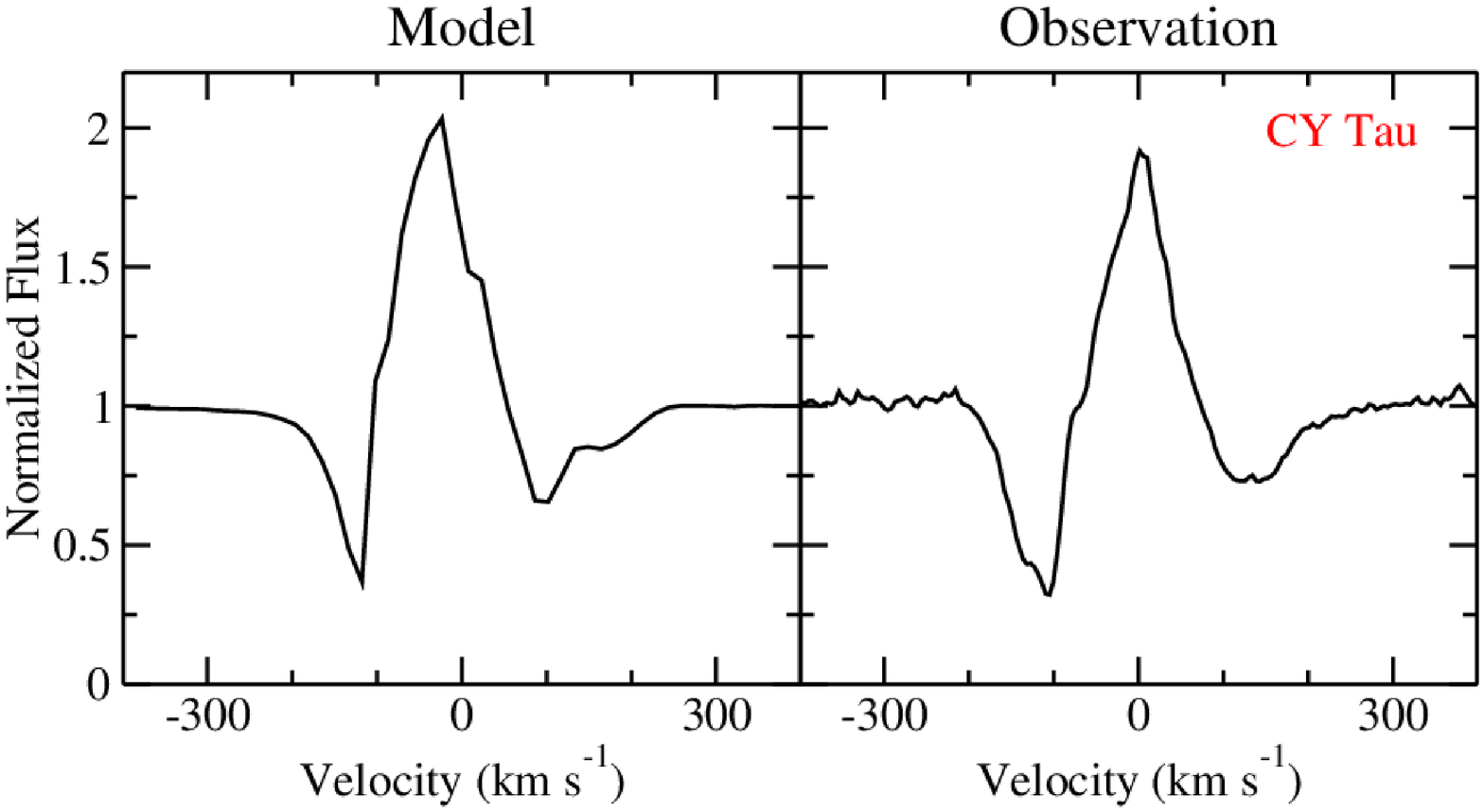} 
  \end{center}
  \vspace{-0.1cm}
  \caption{Summary of the conical wind
    model (\citealt{Romanova:2009}) and a
    qualitative comparison of the model and observed
    \ion{He}{i}~$\lambda$10830 profiles.  \emph{Upper panel}: the poloidal
    matter flux map is over-plotted with the poloidal
    component of the velocity (\emph{arrows}). The \emph{solid
      contours} represent the magnetic field lines. \emph{Lower panels}: a
    qualitative comparison of the model profile of
    \ion{He}{i}~$\lambda$10830 with the observation of
    the CTTS CY~Tau (\citealt{Edwards:2006}). The figures are from \cite{kurosawa12}}
  \label{conical-wind} 
\end{figure*}


To advance our understanding of the formation of the inner winds, we
now apply our radiative transfer models to more realistic winds. For
this purpose, we move our focus back to MHD simulation results. Recent
MHD simulations by \citet{Romanova:2009} have shown a new type of
outflow configuration, so called ``the conical wind'' (the upper panel
in Fig.~\ref{conical-wind}), which is formed when a large scale
stellar dipole magnetic field is compressed by the accretion disk into
the X-wind (\citealt{Shu:1994}) configuration.  The outflows occur in
a rather narrow conical shell, and the outflow speed of the matter is
an order of the Keplerian rotation speed near the wind launching
region. Interestingly, the conical wind model does not require the
magnetospheric radius to be very close the corotation radius, unlike
the X-wind model. Similar simulations with extended computational
domain by \citet{Lii:2012} have shown that the conical wind become
collimated at large distance. The model was applied to 
the high mass-accretion rate FU Ori systems in which the wind can be   
strongly collimated (\citealt{Koenigl:2011}). 

We extended the original conical wind model of \citet{Romanova:2009}
to include a well-defined magnetospheric accretion funnel flow which
is essential for modeling the strong optical and near-infrared
emission lines in CTTSs.  We achieved this by using a slightly
stronger magnetic filed and a lower mass-accretion rate to reduced the
compression of magnetosphere by the disk, hence to form a larger
magnetospheric funnel flow. The resulting flow geometry is shown in
Fig.~\ref{conical-wind} (the upper panel).  We then used the density
and velocity field from the MHD simulation, after it reached a
semi-steady state, in our radiative transfer calculations to predict
hydrogen and helium line profiles.  The line profiles were computed
with various combinations of X-ray fluxes (important for photoionizing
\ion{He}{i}), gas temperatures and inclination angles.  A large
variety of line profile morphology is found, and many of the model
profiles are very similar to those found in observations. An example
of \ion{He}{i}~$\lambda$10830 line profile calculation along with the
observation of the CTTS CY~Tau (from \citealt{Edwards:2006}) is shown
in Fig.~\ref{conical-wind} (the lower panels).  More examples can be
found in \citet{kurosawa12}.  As one can see in
Fig.~\ref{conical-wind}, the conical wind model can well reproduce the
relatively narrow and low-velocity blueshifted wind absorption
component in the observed line profile of \ion{He}{i}~$\lambda$10830.
It also reproduces the redshifted absorption component, which is
caused by the infalling material in the accretion funnel flow.  In
summary, based on our line profile models, we found that the conical
wind model is a very plausible outflow model for a system which shows
a narrow and low-velocity blueshifted absorption component in
\ion{He}{i}~$\lambda$10830, and it is an important alternative model
to a disk wind model which can also form the narrow and low-velocity
blueshifted absorption component, as we have shown in
Sec.~\ref{subsec:simple-wind}.

\section{Summary}
\label{sec:summary}

Recent observations suggest that the geometry of the magnetospheric
accretion may significantly deviate from an axisymmetry for some of
CTTSs. We have briefly reviewed the 3-D MHD simulations which
considered the accretion onto CTTSs with inclined magnetic multipoles
(e.g., a dipole and an octupole).  Non-axisymmetric accretion funnel flows
are naturally found in those simulations.  Using the output from the
simulations in our radiative transfer models,  observable
quantities such as line profiles and light curves are computed. The
predictions can be readily compared with observations.  While the
periodic/regular variability behaviors seen in observations of CTTSs
can be well reproduced by non-axisymmetric accretion models that
are in a steady state (Sec.~\ref{subsec:V2129Oph}), the irregular
variability behaviors seen in some of CTTSs could be explained by the
magnetospheric accretion that is in unstable regime due the magnetic
Rayleigh-Taylor (R-T) instability (Sec.~\ref{subsec:instab}). In
addition to the irregular light curves and line variability, the
unstable accretion due to the R-T instability would produce a variable
but rather persistent redshifted absorption component in higher Balmer
lines (e.g., H$\gamma$ and H$\delta$) and in some near-infrared
hydrogen lines such as Pa$\beta$ and Br$\gamma$, due to the presence
of many accretion streams caused by the instability.

We have also investigated the possible origins of the inner winds in
CTTSs using the line profile models. Using the simple kinematic wind
models in our radiative transfer models
(Sec.~\ref{subsec:simple-wind}), we confirmed the earlier finding
by \citet{Edwards:2006} who suggested that there are two types of
inner winds of CTTSs. We confirmed that the relatively narrow and
low-velocity blueshifted absorption component in
\ion{He}{i}~$\lambda$10830 can be reproduced by a disk wind model, and
the wide and deep blueshifted absorption component in the same line
can be reproduced by a stellar wind model. For the stellar wind
component to be visible in the absorption component, the inclination
angle of the system must be rather small.  We have also tested the
plausibility of the conical wind solution found in the recent MHD
simulations by \citet{Romanova:2009} by using the simulation results
in our radiative transfer models (Sec.~\ref{subsec:conical-wind}). We
found the conical wind model can also reproduce the relatively narrow
and low-velocity blueshifted absorption component in
\ion{He}{i}~$\lambda$10830 seen in some CTTSs.  Finally, high
resolution spectroscopic data and radiative transfer models are
invaluable for testing different outflow and inflow scenarios since a
direct imaging of the inner wind launching regions is still
unattainable for most of the CTTSs.

\section*{Acknowledgment}
We thank the conference organizers for the excellent meeting. RK thanks
Silvia Alencar, Suzan Edwards and Tim Harries for their support and
discussions.  Resources supporting this work were provided by the NASA
High-End Computing (HEC) Program through the NASA Advanced
Supercomputing (NAS) Division at Ames Research Center and the NASA
Center for Computational Sciences (NCCS) at Goddard Space Flight
Center.  The research was supported by NASA grants NNX11AF33G and NSF
grant AST-1211318.

\end{document}